\documentclass{article}
\begin{document}

\title{Cosmological Redshift. An Information Mechanics Perspective}

\author{Frederick W. Kantor$^1$}

\date{}

\maketitle

\footnotetext[1]{523 West 112 Street, New York, NY 10025-1619, USA;
email: kantor@mail.execnet.com ; http://w3.execnet.com/kantor .}

\begin{abstract}
A photon's observed wavelength tells an astronomical detector about the
amount of position information obtained by observing that photon.  This
amount of position information may depend on time in a way which, to first
order over distances small compared to the size of the universe, might
account for the cosmological redshift.

\end{abstract}

\qquad \qquad (in press, \textit{International Journal of Theoretical Physics})

\parskip 4 pt

Interpretation of cosmological redshift as due to rapid recession of
distant light sources has relied on a classical-mechanical picture (Hubble,
1936a, b). Here, we ask about information received in detecting a photon
emitted long ago and far away.  Specifically, the photon's observed
wavelength tells the detector about the amount of position information
accessed by such detection.  This amount of position information may depend
on time in a way which, to first order over distances small compared to the
size of the universe, might account for the cosmological redshift.

In Kantor's information mechanics (\textbf{IM}) (Kantor, 1977; hereinafter 
cited as \textbf{\textit{IM}}), total accessibility of information $I_{\bf 
U}$ (\textbf{\textit{IM}}, p.155) in universe {\textbf U} is finite; there is 
a constant longest possible length $\lambda_1$ (\textbf{\textit{IM}}, p.188); 
\textquotedblleft radius\textquotedblright\ $R_{\bf U}$ of \textbf{U} is 
defined as \mbox{$\lambda_1 /(2 \pi)$} (\textbf{\textit{IM}}, p.212).  The 
amount of information indicated by a photon's wavelength is $\lambda_1$ 
divided by the photon's wavelength (\textbf{\textit{IM}}; Kantor, 1997).

Quantum mechanics (\textbf{QM}) appears in restricted form as a special case 
in \textbf{IM}.  The picture discussed here was reached via \textbf{IM}; it 
is expressed here using terms which may be more familiar from \textbf{QM}.

\subsection*{Position}
\parskip 0 pt
Consider an object in finite universe \textbf{U}, with mass much less than 
the total mass in \textbf{U}, designating a position much smaller than 
\textbf{U}, with de~Broglie wavelength $h/(mv)$ (Planck's constant $h$, mass 
$m$, velocity $v$). For that object to exist in \textbf{U}, its de Broglie 
wavelength would fit in at least once; $v$ would not be zero, although least 
motion so implied might not have any definite direction.

If a photon were emitted substantially isotropically from a source
substantially at rest compared to the average position of distant matter in
the universe, one might say in \textbf{QM} terms that the concept of its
having a direction of propogation would not take on meaning until after an
observation was made. That photon's initial wavelength $\lambda_0$ would
represent a sufficient amount of position information to designate a region
of size \mbox{$\lambda_0 / (2 \pi)$} in \textbf{U}.  While the spherical
shell where the photon could be detected would rapidly expand, the photon's
focal spot would remain at the shell's center, very nearly at rest.

From outside such an expanding not-yet-detected photon shell, consider such a
shell as an object in \textbf{U}: because $R_{\bf U}$ is finite, it would
seem that, unless an extra assumption were added saying that the center of an
expanding photon shell could be exactly stationary, for the shell to exist in
\textbf{U} the center of that object \textemdash\ the expanding shell
\textemdash\ would have nonzero motion.  With such (tiny) motion spreading
the initially designatable region in \textbf{U}, absent significant
electromagnetic interaction between the expanding shell and interstellar
matter, the photon's amount of position information accessible by an
astronomical detector would decrease after the photon was emitted and before
it was detected.

During expansion of the not-yet-detected photon shell, the photon would not
be detected within the part of \textbf{U} through which the shell had already
expanded.  For this reason, one might make a causality argument that the de
Broglie wavelength for that expanding shell fit into the portion of
\textbf{U} outside the shell; this would give rise to an increasing rate of
spread of the focal spot before the photon was detected at larger distance.

Also, re increased rate of focal spot spread with larger distance before
photon detection, one might consider whether a reduction of position
information might in effect appear as a reduction of the mass of the photon
shell object, requiring more rapid motion of the shell's center for the
shell's de~Broglie wavelength to fit in.

Although consideration of increased rate of motion of the center of a
not-yet-detected photon shell would support relatively large redshift for
photon detection at sufficiently large distance, such contribution(s) would
be relatively small for distances sufficiently small compared to the size of
\textbf{U}.  That lets us consider the behavior at sufficiently small
distances to obtain an approximate value to \textquotedblleft first
order\textquotedblright\ for the magnitude of this redshift effect.

\subsection*{Redshift}
Consider a visible photon (e.g., \mbox{$\sim \!500$ nm} wavelength) emitted,
with initial wavelength $\lambda_0$, substantially isotropically from a
source substantially at rest compared to the average position of distant
matter in the universe.  The initial energy of the photon, expressed in units
of mass $m$, would be \mbox{$h/(\lambda_0 c)$}, where $c$ denotes speed of
light.  The initial de~Broglie wavelength, $h/(mv)$, of the not-yet-detected
photon shell would thus be \mbox{$\lambda_0 c / v$}.  Let \mbox{$\lambda_0
\ll d \ll R_{\bf U}$}, where $d$ is the distance from source to detector.
Requiring that the de~Broglie wavelength fit once into \mbox{$2 \pi R_{\bf
U}$}, and solving for least motion speed $v_1$, the first-order term for the
least possible rate of motion of the center of the not-yet-detected photon
shell would be
\begin{equation}
\qquad \qquad \qquad \qquad
v_1 \approx \frac{\lambda_0 c}{2 \pi R_{\bf U}} + \cdots
\qquad \qquad \qquad \qquad
(\lambda_0 \ll d \ll R_{\bf U})
\quad
\end{equation}
During the time \mbox{$t \approx d/c$} between when the photon was emitted
and when it was detected, the original size of the region which the photon
was able to designate, \mbox{$\lambda_0 / (2 \pi )$}, would have grown (least
first-order term) by an amount \mbox{$\approx v_1 t$}.  The new wavelength
$\lambda(d)$ seen at distance $d$, corresponding to representation of the
amount of position information associated with designating this slightly
larger region, would then be given to first order by
\begin{equation}
\ \qquad \qquad \qquad
\frac{\lambda (d)}{2 \pi} \approx \frac{\lambda_0}{2 \pi} +
\frac{\lambda_0 c}{2 \pi R_{\bf U}} \frac{d}{c} + \cdots
\ \ \qquad \qquad \qquad
(\lambda_0 \ll d \ll R_{\bf U})
\quad
\end{equation}
Canceling, and collecting terms, this would give as a least first-order
effect of distance on wavelength
\begin{equation}
\qquad \qquad \qquad
\lambda(d) \approx \lambda_0 (1 + d/R_{\bf U} + \cdots)
\quad \qquad \qquad \qquad
(\lambda_0 \ll d \ll R_{\bf U})
\quad
\end{equation}
For distances much smaller than the size of the universe, this first-order
approximation for cosmological redshift with distance appears similar in
size to that of the older picture, here writing redshift directly in terms
of distance.

\subsection*{Apparent time dilation of remotely observed event}
For an event occurring in much less time than it takes for light from there
to reach the observer, consider the above photon redshift with regard to
frequency components of the event's detected signature.  Noting that the
wavelength appears outside the parentheses in (3), observation of the
combined result would thus appear stretched in time by about the same ratio
as the increase in photon wavelengths.

\subsection*{Discussion}
What does this mean?  In physics, a theory is only required to deal with
information accessible to an observer.  Here, consideration of the amount of
position information accessed by an observer might seem able to account to
first order for the redshift due to distance seen in light received from
nearby sources.  From this point of view it would seem that, without having
to specify a particular value for $R_{\bf U}$, the assumption that the
cosmological redshift be caused by rapid recession of the light source, might
not be unique.

\subsection*{REFERENCES}
\noindent
Hubble, E. (1936a). \textit{Proceedings of the National Academy of Sciences
USA}, \textbf{22}, 621.
\parskip 3 pt

\noindent
Hubble, E. (1936b). \textit{The Realm of the Nebulae}, Yale University
Press, New Haven, Connecticut.

\noindent
Kantor, F.W. (1977). \textit{Information Mechanics}, Wiley, New York.

\noindent
Kantor, F.W. (1997). \textit{International Journal of Theoretical Physics},
\textbf{36}(6), 1317 \textendash\ 1319;
\mbox{ http://w3.execnet.com/kantor/pm00.htm} (web citation not in journal).

\end{document}